\documentclass[pra,twocolumn,a4paper,showpacs]{revtex4}
\usepackage{latexsym}
\usepackage{amsmath}
\usepackage{amsfonts}
\usepackage{graphicx}
\usepackage{exscale}
\usepackage{amssymb}
\usepackage{dsfont}
\usepackage{natbib}

\newcommand{\ket}[1]{\ensuremath{|#1\rangle}}
\newcommand{\bra}[1]{\ensuremath{\langle #1 |}}

\newcommand{\ve}{\varepsilon}
\newcommand{\eps}{\epsilon}
\newcommand{\vro}{\varrho}
\newcommand{\be}{\begin{equation}}
\newcommand{\ee}{\end{equation}}
\newcommand{\ba}{\begin{eqnarray}}
\newcommand{\ea}{\end{eqnarray}}
\newcommand{\mc}[1]{\ensuremath{\mathcal{#1}}}
\newcommand{\mf}[1]{\boldsymbol{#1}}
\newcommand{\mfi}[1]{\boldsymbol{#1}}
\newcommand{\Sp}[2]{S_{#1 \,+}^{  (#2)}}
\newcommand{\Sm}[2]{S_{#1 \,-}^{  (#2)}}
\newcommand{\TensorRe}{ \overset{\leftrightarrow}{\chi}_{\text{re}}}
\newcommand{\TensorIm}{ \overset{\leftrightarrow}{\chi}_{\text{im}}}
\newcommand{\rot}[3][]{\mc{R}_{#2}^{#1}(#3)}

\newcommand{\meanb}[1]{\ensuremath{ \big\langle\,#1\, \big\rangle}}

\begin{document}

\title{Breakdown of the few-level approximation in collective systems}
\author{M. \surname{Kiffner}}
\email{martin.kiffner@mpi-hd.mpg.de}

\author{J. \surname{Evers}}
\email{joerg.evers@mpi-hd.mpg.de}

\author{C. H. \surname{Keitel}}
\email{keitel@mpi-hd.mpg.de}

\affiliation{Max-Planck-Institut f\"ur Kernphysik, 
Saupfercheckweg 1, 69117 Heidelberg, Germany}


\date{\today}

\begin{abstract} 
The validity of the few-level approximation in dipole-dipole
interacting collective systems is discussed. 
As example system, we study the archetype case of 
two dipole-dipole interacting atoms,
each modelled by two complete sets
of angular momentum multiplets.
We establish the breakdown of the few-level approximation by
first proving the intuitive result that the dipole-dipole induced energy 
shifts between collective two-atom states depend on the 
length of the vector connecting the atoms, but not on its orientation,
if complete and degenerate multiplets are considered. 
A careful analysis of our findings reveals that the simplification of the 
atomic level scheme by artificially omitting Zeeman sublevels in a few-level
approximation generally leads to incorrect predictions. 
We find that this breakdown can be traced back to the dipole-dipole
coupling of transitions with orthogonal dipole moments.
Our interpretation enables us to identify special geometries in which 
partial few-level approximations to two- or three-level systems are valid.
\end{abstract}

\maketitle

\section{\label{sec:intro}Introduction}

The theoretical analysis of any non-trivial physical 
problem typically requires the use of approximations.
A key approximation facilitated in most areas of physics 
reduces the complete configuration space of the system of 
interest to a smaller set of relevant system states. 
In the theoretical description of atom-field interactions,  
the essential state
approximation entails neglecting most of the bound and continuum  
atomic states~\cite{zubairy:qo,ficek:int,agarwal:qst2}. 
The seminal Jaynes-Cummings-Model~\cite{jaynes:63}
takes this reduction to the extreme in that only two
atomic states are retained. 
Obviously, it is essential to in detail explore the
validity range of this reduction of the configuration space. 
The few-level approximation usually leads to 
theoretical predictions that are well verified 
experimentally~\cite{zubairy:qo,ficek:int}, and is generally 
considered as understood for single-atom systems.
It fails, however, to reproduce results of quantum 
electrodynamics, where in general all possible intermediate
atomic states need to be considered in order to obtain
quantitatively correct results~\cite{lamb:87}.
The situation becomes even less clear in collective systems, 
where the individual constituents interact via the dipole-dipole 
interaction, despite the relevance of collectivity to many
areas of physics. Examples for such systems can be found 
in ultracold quantum gases~\cite{ref6to8}, trapped 
atoms~\cite{refexp,kastel:05},
or solid state systems~\cite{hettich:02,barnes:05}, with applications, 
e.g., in quantum information 
theory~\cite{ref14to17}.  

\begin{figure}[t]
\includegraphics[scale=1]{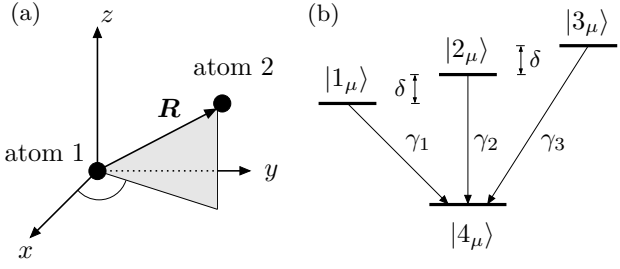}
\caption{\label{picture1}
(a) The system of interest 
 is comprised of two identical atoms that are located 
at $\mf{r}_1$ and $\mf{r}_2$, respectively. 
$\mf{R}=\mf{r}_2 -\mf{r}_1$  is the relative position of atom 2 with respect to atom 1.  
(b) Level structure of atom $\mu\in\{1,2\}$ which we employ to illustrate our results.  
The ground state  is a $S_0$ singlet state, and the three excited levels 
are Zeeman sublevels of a $P_1$ triplet. 
$\delta$ is the frequency splitting of the upper levels.}
\end{figure}

Therefore, we discuss the validity of the few-level  
approximation in dipole-dipole interacting collective systems.
For this, we study the archetype case of two dipole-dipole 
interacting atoms, see Fig.~\ref{picture1}(a).  
Experiments of this type have become possible
recently~\cite{refexp,hettich:02}. 
In order to remain general, each atom is modelled by complete sets
of angular momentum multiplets, as shown in 
Fig.~\ref{picture1}(b).
We find that the few-level approximation in general leads to 
incorrect predictions if it is applied to the magnetic sublevels 
of this system. 
For this, we first establish a general statement about the 
system behavior  under rotations of the atomic separation vector $\mf{R}$.  
As a first conclusion from this result, we derive
the intuitive outcome that  
the dipole-dipole induced energy shifts between 
collective two-atom states are invariant under rotations of the 
separation vector $\mf{R}$. 
This result can only be established if complete  and 
degenerate multiplets are considered and  
dipole-dipole interactions between orthogonal
transition dipole moments are included in the 
analysis.
On the contrary, the artificial omission of  any of the 
Zeeman sublevels of a multiplet leads to 
a spurious dependence of the energy shifts on the orientation,
and thus to incorrect predictions.

For example, if in the well-known two-level approximation
only one excited state $\ket{e}$ and the ground state 
$\ket{g}$ are retained, then we recover the position-dependent 
energy splitting between the entangled two-particle states 
$(\ket{e,g}\pm \ket{g,e})/\sqrt{2}$ that has previously
been reported for a pair of two-level 
systems~\cite{agarwal:qst2,ficek:int}. 
This geometry-dependence is at odds with the rotational invariance
of the collective energy splitting expected for the 
degenerate system with all Zeeman sublevels.
We thus conclude that  the few-level approximation in general 
cannot be applied to this system.

Our results can be generalized to more complex 
angular momentum multiplets.

%%%%%%%%%%%%%%%%%%%%%%%%%%%%%
%%%%%%%%%%%%%%%%%%%%%%%%%%%%%%%%%%%%

\section{\label{sec:model}The Model}

We describe each atom by  a 
$S_0\leftrightarrow P_1$ transition shown in Fig.~\ref{picture1}(b) 
that can be found, e.g., in ${}^{40}$Ca atoms. 
We choose the $z$ axis as the quantization axis,
which is distinguished by an external magnetic 
field that induces a Zeeman splitting 
$\delta$ of the excited states.  
The orientation of $\mf{R}$ 
is defined relative to this
quantization axis.
We begin with  the introduction of the master equation which governs the 
atomic evolution of the system shown in Fig.~\ref{picture1}. 
The internal state  $\ket{i_{\mu}}$ of atom $\mu$  
 is an eigenstate of  $J_z^{(\mu)}$, where  $\mf{J}^{(\mu)}$ is  the angular momentum 
operator of atom $\mu$ ($\mu\in\{1,2\}$). In particular, 
the $P_1$ multiplet with $J=1$ corresponds to the excited states 
$\ket{1_{\mu}}$, $\ket{2_{\mu}}$ and  $\ket{3_{\mu}}$  with  
magnetic quantum numbers  $m=-1,\,0$ and 1, respectively, 
and the $S_0$ state is the ground state $\ket{4_{\mu}}$ with $J=m=0$.
The   raising and lowering operators on the 
\mbox{$\ket{4_{\mu}}\leftrightarrow\ket{i_{\mu}}$} 
transition  of atom $\mu$ are ($i\in\{1,2,3\}$)
\be
\label{transition-ops}
\Sp{i}{\mu} = \ket{i_\mu}\bra{4_{\mu}}  \quad\text{and}\quad  \Sm{i}{\mu} 
=\ket{4_{\mu}} \bra{i_\mu}\,.
\ee
The total system Hamiltonian  for the two atoms and the radiation field  is 
$
H= H_{\text{A}} +H_{\text{F}} + V \,,
$
where 
\begin{align}
H_{\text{A}}   = &  \hbar \sum\limits_{i=1}^3  \sum\limits_{\mu=1}^{2}\omega_i \, \Sp{i}{\mu} \Sm{i}{\mu}\,,
\quad H_{\text{F}} =  \sum\limits_{\mfi{k}s}\hbar \omega_k a_{\mfi{k}s}^{\dagger} a_{\mfi{k}s}\,, \notag\\ 
V  = & 
 - \mf{\hat{d}}^{(1)}\cdot \mf{\hat{E}}(\mf{r}_{1})- \mf{\hat{d}}^{(2)}\cdot \mf{\hat{E}}(\mf{r}_{2})\, .
\label{H}
\end{align}
In these equations, $H_{\text{A}}$ describes the free evolution of the two identical atoms, 
$\hbar\omega_i$ is the energy of state $\ket{i_{\mu}}$  and  
we choose  $\hbar\omega_4=0$.  
$H_{\text{F}}$ is the Hamiltonian of the  vacuum field  and $V$ 
describes the interaction of the atoms with the vacuum modes in dipole approximation.
The electric field operator $\mf{\hat{E}}$ is defined as 
\be
\mf{\hat{E}}(\mf{r})=i\sum\limits_{\mfi{k}s}
\sqrt{\frac{\hbar\omega_k}{2\ve_0 v}}\mf{\eps}_{\mfi{k}s}e^{i\mfi{k}\cdot\mfi{r}}  a_{\mfi{k}s} +\text{H.c.}\,,
\ee
where $a_{\mfi{k}s}$ ($a_{\mfi{k}s}^{\dagger}$) are the annihilation (creation) operators that 
correspond to a field mode with wave vector $\mf{k}$, polarization $\mf{\eps}_{\mfi{k}s}$ 
and frequency $\omega_k$, and $v$ denotes the quantization volume. 
The electric-dipole moment operator of atom $\mu$  is a vector operator   with respect to  
the angular momentum operator $\mf{J}^{(\mu)}$ of atom $\mu$ and 
reads
\be
\mf{\hat{d}}^{(\mu)}=\sum\limits_{i=1}^{3}\big[\mf{d}_{i}  \Sp{i}{\mu} +\text{H.c.}\big]\,. 
\label{dipole_op}
\ee
We determine the dipole moments $\mf{d}_i =   \bra{i}\mf{\hat{d}}\ket{4}$ 
via the Wigner-Eckart theorem~\cite{sakurai:mqm} and find
\begin{subequations}
\label{dipoles}
\begin{align}
\mf{d}_1  &=    \mc{D}\,\mf{\eps}^{(+)}\,, \qquad
\mf{d}_3   = - \mc{D}\,\mf{\eps}^{(-)}\,,\\
\mf{d}_2  &=   \mc{D}\,\mf{e}_z
 \,, 
\end{align}
\end{subequations}
where $\mc{D}$ is the reduced dipole matrix element 
and the circular polarization vectors are $\mf{\eps}^{(\pm)}=(\mf{e}_x \pm i \mf{e}_y)/\sqrt{2}$.  

We now adapt the standard derivation of a master 
equation~\cite{zubairy:qo,ficek:int,agarwal:qst2} to our multilevel
system. For this, 
we assume that the radiation 
field is initially in the vacuum state 
denoted by  $\vro_{\text{F}}$ and 
suppose that the total density operator factorizes into a  product of $\vro_{\text{F}}$ and 
the atomic density operator $\vro$ at $t=0$. 
The master equation  for the reduced atomic density operator  
in   Born approximation    then takes the form
\begin{align}
\partial_t\vro=&-\frac{i}{\hbar}[H_{\text{A}},\vro] \label{master_general} \\
& -\frac{1}{\hbar^2}\int\limits_0^{t}
d\tau \text{Tr}_{\text{F}}\Big(\big[V, U(\tau)\big[V,\vro_{\text{F}}\vro(t-\tau)\big]U^{\dagger}(\tau)\big]\Big)
\,,\notag
\end{align}
where 
$U(\tau) = \exp[-i(H_{\text{A}} + H_{\text{F}})\tau/\hbar]$ 
and $\text{Tr}_{\text{F}}()$ denotes the trace over the vacuum modes. 
We evaluate the integral in Eq.~(\ref{master_general}) 
in Markov-approximation~\cite{zubairy:qo} 
and ignore all terms associated with the Lamb shift of the atomic levels.
In addition, we employ the rotating-wave approximation 
and neglect anti-resonant terms that are proportional 
to $\Sp{i}{\mu}\Sp{j}{\nu}$ and $\Sm{i}{\mu}\Sm{j}{\nu}$.  
We finally obtain 
\be
\partial_t \vro  =  -\frac{i}{\hbar} [H_{\text{A}} , \vro] -\frac{i}{\hbar} [H_{\Omega} , \vro] + \mc{L}_{\gamma}\vro \,.
\label{master}
\ee
In this equation, the Hamiltonian $H_{\Omega}$ 
describes the coherent part of the 
dipole-dipole interaction and reads 
\be
 H_{\Omega}   =    
 -\hbar  \sum\limits_{i,j=1}^3  \left\{ \Omega_{ij} \Sp{i}{2} \Sm{j}{1}  \,+\,\text{H.c.} \right\}   \,.
\label{H_Omega}
\ee
The coefficients $\Omega_{ij}$ are defined as~\cite{agarwal:01,evers:06}
\be
\Omega_{ij}  =\frac{1}{\hbar} \left[\mf{d}_i^{\text{T}} \;
\TensorRe(\mf{R}) \;\mf{d}_j^* \right] \,, 
\label{Omega} 
\ee
and the tensor $\TensorRe$ is the real part of the tensor 
 $\overset{\leftrightarrow}{\chi}$  
whose components $\overset{\leftrightarrow}{\chi}_{kl}$ for $k,l \in\{1,2,3\}$ 
are given by 
\be
\overset{\leftrightarrow}{\chi}_{kl}(\mf{R})  =\frac{k_0^3}{4\pi\ve_0}\left [  g_1(\eta)\,\delta_{kl}
 - g_2(\eta)\,\frac{\mf{R}_{k}\mf{R}_{l }}{R^2} 
\right ]\,e^{i \eta}\,. 
\label{chi}
\ee
Here the vector  $\mf{R}$ denotes the  relative coordinates 
of \mbox{atom 2} with respect to \mbox{atom 1}  [see Fig.~\ref{picture1}(a)], $\eta=k_0 R$ and 
  $g_1 =( \eta^{-1}+ i \eta^{-2} - \eta^{-3})$,  
$g_2  = ( \eta^{-1}+ 3i\eta^{-2} - 3\eta^{-3} )$. 
In the derivation of Eq.~(\ref{chi}), the three transition frequencies $\omega_1$, $\omega_2$ and 
$\omega_3$ have been 
approximated  by their mean value $\omega_0=c k_0$ ($c$: speed of light)~\cite{remark:1}.  
This is justified since the  Zeeman splitting $\delta$ is  
much smaller  than the optical transition frequencies $\omega_i$.  
%%%%%%%%%%%%%%%%%%%%%%%%

The last term in Eq.~(\ref{master}) accounts for spontaneous emission and reads 
\begin{align}
&\mc{L}_{\gamma} \vro  =           
 -\!\!\sum\limits_{\mu=1}^2\!\sum\limits_{i=1}^3 
\gamma_i \!\left( \Sp{i}{\mu} \Sm{i}{\mu} \vro + \vro \Sp{i}{\mu}\Sm{i}{\mu} - 2 \Sm{i}{\mu} \vro \Sp{i}{\mu}
\right)  \notag \\
& -  \! \!\sum\limits_{\genfrac{}{}{0pt}{2}{\mu,\nu =1}{\mu\not=\nu}}^2
 \sum\limits_{i,j =1}^3  
\Gamma_{ij} \left(
\Sp{i}{\mu} \Sm{j}{\nu} \vro + \vro \Sp{i}{\mu}\Sm{j}{\nu} - 2 \Sm{j}{\nu}\vro \Sp{i}{\mu}\right)
   \,.  \label{Lgamma}
\end{align}
The total decay rate of the excited state $\ket{i}$ of each of the atoms is 
given by $2\gamma_i$, where 
$
\gamma_i = |\mf{d}_i|^2 \omega_0^3/(6\pi\eps_0\hbar c^3)=\gamma 
$
and we again employed the approximation \mbox{$\omega_i \approx \omega_0$}.
The collective decay rates $\Gamma_{ij}$ result from the vacuum-mediated dipole-dipole coupling 
between the two atoms and are determined by 
\be
\Gamma_{ij} = 
\frac{1}{\hbar} 
\left[\mf{d}_i^{\text{T}} \;  \TensorIm(\mf{R})\; \mf{d}_j^* \right] 
\, , 
 \label{Gamma}
\ee
where $\TensorIm=\text{Im}\overset{\leftrightarrow}{\chi}$ is the imaginary 
part of the tensor $\overset{\leftrightarrow}{\chi}$. 
Note that the cross terms ($i\neq j$) in Eqs.~(\ref{Omega}) and (\ref{Gamma})
represent couplings between transitions with orthogonal dipole moments. 
If  the master equation~(\ref{master}) is transformed into 
the interaction picture with respect to $H_{\text{A}}$, 
terms proportional to these cross terms 
rotate at   frequencies $\pm\delta$ or $\pm 2\delta$. 
It follows that the parameters $\Omega_{ij}$ and $\Gamma_{ij}$ 
are negligible if the level splitting $\delta$ is 
large, i.e. $|\delta| \gg |\Omega_{ij}|, |\Gamma_{ij}|$ ($i\neq j$). 

%%%%%%%%%%%%%%%%%%%%%%%%

Next we provide explicit  expressions for the coupling constants $\Omega_{ij}$ and 
the decay rates $\Gamma_{ij}$ in Eqs.~(\ref{Omega}) and (\ref{Gamma}), respectively. 
For this, it is convenient to express the  relative position of the two atoms in spherical coordinates,  
\be
\label{coord}
\mf{R} =R \, (\sin\theta\cos\phi,\,\sin\theta\sin\phi,\,\cos\theta)\,. 
\ee
Together with Eqs.~(\ref{chi}) and ~(\ref{dipoles}) we obtain 
\begin{align}
\Omega_{31} &= \gamma\frac{3}{4\eta^3}\left[\left(\eta^2-3\right)\cos\eta-3\eta\sin\eta\right]\sin^2\theta e^{-2 i \phi}\,,
\notag \\[0.1cm]
\Omega_{11} &  =  3\frac{\gamma}{8\eta^3}\left[\left(3\eta^2-1+\left(\eta^2-3\right)\cos2\theta\right)\cos\eta\right. 
\hspace*{1cm} \notag\\[0.1cm]
& \left.\hspace*{3.5cm} -\eta\left(1+3\cos2\theta\right)\sin\eta\right]\,, \notag \\[0.1cm]
\Omega_{21}    &=     -\sqrt{2} \cot\theta\;\Omega_{31} e^{i\phi}        \,,     \notag \\[0.1cm]
 \Omega_{22}    &=            \Omega_{11} - (2 \cot^2\theta -1) \Omega_{31} e^{2i\phi} \,,  \notag \\[0.2cm]
 \Omega_{32}    &=-\Omega_{21}\,,\quad \Omega_{33}=\Omega_{11}\,,
 \label{OmegaExplicit}
 \end{align}
and the collective decay rates evaluate to
\begin{align}
\Gamma_{31}  & =  \gamma\frac{3}{4\eta^3}\left[\left(\eta^2-3\right)\sin\eta + 3\eta\cos\eta\right]\sin^2\theta e^{-2 i \phi}
\,,\notag \\[0.1cm]
\Gamma_{11} &  =  3\frac{\gamma}{8\eta^3}\left[\left(3\eta^2-1+\left(\eta^2-3\right)\cos2\theta\right)\sin\eta\right. 
\hspace*{1cm} \notag\\[0.1cm]
 &\left.\hspace*{3.5cm} +\eta\left(1+3\cos2\theta\right)\cos\eta\right] \,, \notag\\[0.1cm]
\Gamma_{21} &=      -\sqrt{2} \cot\theta\;\Gamma_{31}e^{i\phi}       \,,       \notag \\[0.1cm]
\Gamma_{22}&=            \Gamma_{11} - (2 \cot^2\theta -1) \Gamma_{31}   e^{2 i\phi} \,,  \notag \\[0.2cm]
 \Gamma_{32} & =-\Gamma_{21}\,,\quad \Gamma_{33}=\Gamma_{11}\,.
 \label{GammaExplicit}
 \end{align}
A numerical study of these coupling terms can be found in~\cite{dfs}.

%%%%%%%%%%%%%%%%%%%%%%%%%%%%%%%%%%%%%%%%%%%%%%%%%%%%%%%%%
\begin{figure}[t]
\includegraphics[width=8cm]{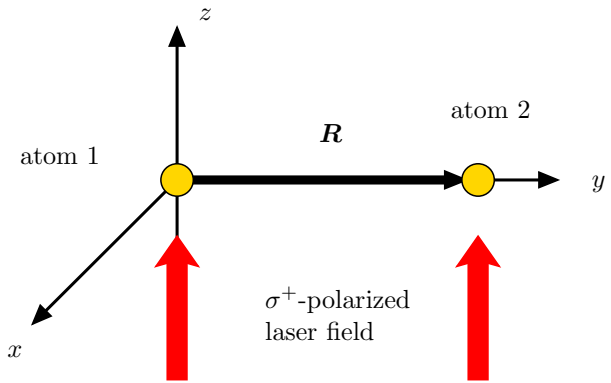}
\caption{\label{fig:example-setup}(Color online)
Setup considered
in Section~\ref{sec:example}, which provides an
illustration of the physical mechanisms responsible
for the breakdown of the few-level approximation.
In this example, an external
laser field is used for the
sake of illustration. Our main results starting from
Section~\ref{sec:analysis} do not rely on external
driving fields.}
\end{figure}
%%%%%%%%%%%%%%%%%%%%%%%%%%%%%%%%%%%%%%%%%%%%%%%%%%%%%%%%%

\section{\label{sec:example}Physical motivation}
In the following Section~\ref{sec:analysis}, we will provide
a rigorous treatment of the behavior of our model system
under rotations of the atomic separation vector in order
to study the geometrical properties of the different coupling
terms in the master equation~(\ref{master}).
In order to motivate this analysis, in this Section~\ref{sec:example}, 
we will discuss a simple example for our results.  
This example employs an  external 
laser field driving the atoms, which is used, 
however, only for the
sake of illustration. Our main results starting from
Section~\ref{sec:analysis} will not rely on external
driving fields.

To this end, we consider  the geometrical setup shown in
Fig.~\ref{fig:example-setup}. 
The atoms with internal structure as 
in Fig.~\ref{picture1} are aligned along the $y$ axis, and 
in addition to the model considered so far, 
a $\sigma^+$ polarized laser beam with Rabi frequency 
$\Omega_L$ and frequency $\omega_L$ propagates in 
 $z$-direction. In rotating-wave approximation, the atom-laser 
interaction Hamiltonian  reads
\[
H_L   = \hbar \sum\limits_{\mu=1}^{2} \left\{
\Omega_L \,\Sp{3}{\mu} \,e^{-i\omega_L t}    \,+\,\text{H.c.} \right\}  \,.
\]
The transition operators $\Sp{i}{\mu}$ are defined 
in Eq.~(\ref{transition-ops}). Since the laser polarization 
is $\sigma^+$, it couples only to the transition 
$\ket{3}\leftrightarrow\ket{4}$ in each atom. 
To describe this setup, one might be tempted to
employ the usual few-level approximation,
and thus neglect 
the excited states $\ket{1}$ and $\ket{2}$ in 
each atom, since they are not populated by the 
laser field. If this were correct, 
the seemingly  relevant subsystem would 
be 
\[
S = \text{Span}(\ket{4,4},\,\ket{3,3},\,\ket{3,4},\,\ket{4,3})\,.
\]
However, it is easy to prove that the state space of the two 
atoms  can {\it not}
be reduced to the  subspace $S$, i.e., that the 
few-level approximation cannot be applied in its usual form. 
In order to show this, we include the atom-laser interaction into  the master equation~(\ref{master})  
and transform the resulting master equation in a frame rotating with 
the laser frequency.  This equation is solved numerically with 
the initial condition $\vro(t=0)=\ket{4,4}\bra{4,4}$, i.e. it 
is assumed that both atoms are initially in their ground states. 
%%%%%%%%%%%%%%%%%%%%%%%%%%%%%%%%%%%%%%%%%%%%%%%%%%%%%%%%%
\begin{figure}[t]
\includegraphics[scale=1]{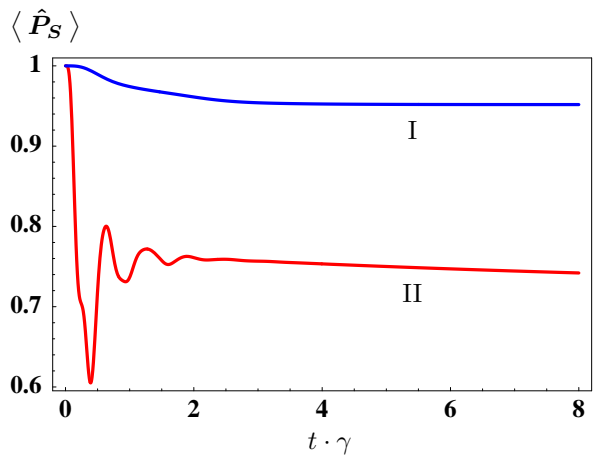}
\caption{\label{fig:example-1}(Color online)
Population in the subspace $S$ obtained by applying
the few-level approximation to the setup in 
Fig.~\ref{fig:example-setup}.
The common parameters are $\theta =\pi/2$, $\phi =\pi/2$, $\delta = 0$. Further, in (I), 
$\Omega_L  =2 \gamma$, 
$R=0.3\,\lambda_0$, and $\Delta = 0.58\,\gamma$, where 
$\Delta = \omega_L - \omega_0$. 
Curve (II) shows the case 
$\Omega_L  = 5.4\, \gamma$,
$R=0.1\,\lambda_0$, and $\Delta = 5.2\,\gamma$.
}
\end{figure}
%%%%%%%%%%%%%%%%%%%%%%%%%%%%%%%%%%%%%%%%%%%%%%%%%%%%%%%%%

Figure~\ref{fig:example-1} shows the total population
confined to the  subspace $S$, 
\begin{equation}
\meanb{\hat{P}_{S}}= \text{Tr}\left[\vro(t)
 \,\hat{P}_{S}\right]\,,
\end{equation}
where ${P}_{S}$ is the projector onto the subspace $S$.  
It can easily be seen that for both sets of parameters,
population is lost from the subspace $S$. Since
all states but the excited states $\ket{1}$
and $\ket{2}$ are contained in $S$,  
it is clear that it 
is not sufficient to take only the excited state $\ket{3}$ 
into account in the usual few-level approximation.

The explanation of this outcome is straightforward. 
According to Eq.~(\ref{H_Omega}), the 
dipole transition $\ket{3}$
$\leftrightarrow\ket{4}$ of one atom is 
coupled by the cross-coupling term $\Omega_{31}$ to 
the $\ket{1}\leftrightarrow\ket{4}$ transition of the 
other atom.  This coupling results in a population of 
state $\ket{1}$, even though the transition dipoles
of the two considered transitions are orthogonal.
Consequently, the dipole-dipole interaction between transitions with 
orthogonal dipole moments will result in the (partial) population of 
the states $\ket{1,1}$, $\ket{1,3}$, $\ket{3,1}$, $\ket{1,4}$, $\ket{4,1}$, although 
none of these states is directly coupled to the  laser field. 
%%%%%%%%%%%%%%%%%%%%%%%%%%%%%%%%%%%%%%%%%%%%%%%%%%%%%%%%%
\begin{figure}[t!]
\includegraphics[scale=1]{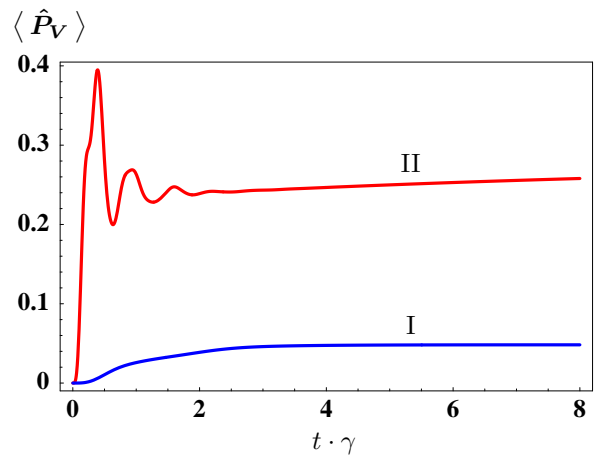}
\caption{\label{fig:example-2}(Color online)
Population of the subspace $V$, which contains all
population which was lost from subspace $S$ in
Fig.~\ref{fig:example-1}, such that the population
in $S+V$ remains unity for all times.
The parameters are as in Fig.~\ref{fig:example-1}.}
\end{figure}
%%%%%%%%%%%%%%%%%%%%%%%%%%%%%%%%%%%%%%%%%%%%%%%%%%%%%%%%%

The numerical verification of these statements is shown
in Figure~\ref{fig:example-2}, 
which depicts the population of the 
subspace 
\[
V = \text{Span}(\ket{1,1},\,
\ket{1,3},\,\ket{3,1},\,\ket{1,4},\,\ket{4,1})\,.
\]
$\hat{P}_V$ is the projector onto the subspace $V$, and 
the parameters are the same as above. 
Note that we have verified that all population is 
contained in the subspace $S+V$, i.e. 
$\meanb{\hat{P}_{S}}+\meanb{\hat{P}_{V}} =1 $ at all times. 

It is important to note that the sufficient subspace $S+V$ still
does not contain all possible states of the two atoms,
because the excited state $\ket{2}$ of each atom
is neglected. The justification for this is that 
in the chosen geometry, the cross-coupling terms 
$\Omega_{21}$, $\Gamma_{21}$ and $\Omega_{32}$, $\Gamma_{32}$ 
vanish such that the transition $\ket{2}\leftrightarrow\ket{4}$ 
of one atom is not coupled to 
the transitions $\ket{1}\leftrightarrow\ket{4}$ and 
$\ket{3}\leftrightarrow\ket{4}$ of 
the other atom, see Eqs.~(\ref{OmegaExplicit}) and (\ref{GammaExplicit}). 
This is important since it demonstrates
that it is also not correct to simply state that all atomic
states have to be taken into account for all parameter 
configurations.

The above example clearly demonstrates that the 
few-level approximation is rendered impossible by 
the coupling terms 
between transitions with {\it orthogonal dipole moments}.  
Therefore, it is the nature of the dipole-dipole coupling 
itself which enforces that generally all Zeeman sublevels have to be taken into account, and not the polarization of the external
laser fields, as one may be tempted to assume in the
usual few-level approximation. 

A physical interpretation for the origin of the
vacuum-induced coupling of transitions with orthogonal dipole moments has been
given in~\cite{evers:06}. In essence, these couplings occur if the 
polarization of a (virtual) photon emitted on one of the transitions in the first atom 
has non-zero projection on different dipole moments of the second atom.
Pictorially, then the second atom cannot measure the polarization
of the photon, and thus has finite probability to absorb it also
on transitions with dipole moments orthogonal to the dipole of the 
emitting transition. 
%%%%%%%%%%%%%%%%%%%%%%%%%%%%%%%%%%%%%%%%%%%%%%%%%%%%%%%%%

\section{\label{sec:analysis}Breakdown of the Few-Level Approximation}

In this section, we return to our original 
setup in Fig.~\ref{picture1}, and thus drop
the external driving fields employed in
Sec.~\ref{sec:example}.
We first derive a general statement 
about the behavior of the master 
equation~(\ref{master}) under rotations 
of the separation vector $\mf{R}$. 
On first sight, we will prove an obvious
result: In the absence of external
fields but the isotropic vacuum, there is no 
distinguished direction in space. Thus one expects
the eigenenergies of the system to be invariant
under rotations of $\mf{R}$, and this is indeed what
we find. But despite its intuitiveness, this statement
needs proof, and the discussion of the proof 
and its assumptions 
will provide the theoretical foundation for our 
central results and physical interpretations 
in the following sections.

\subsection{\label{sec:theorem}Central theorem} 
%%%%%%%%%%%%%%%%%%
% statement
%%%%%%%%%%%%%%%%%
In addition to a given relative position $\mf{R}$ 
of the two atoms, we consider a different geometrical setup 
where the separation vector $\mf{P}$ is obtained from $\mf{R}$ by a 
rotation, $\mf{P}=\rot{\mf{u}}{\alpha} \mf{R}$. Here,
$\rot{u}{\alpha}$ is an orthogonal $3\times3$ matrix 
that describes a rotation in the three-dimensional real 
vector space $\Bbb{R}^3$ around the axis $\mf{u}$ by an angle $\alpha$. 
Our aim is to show that there exists a unitary operator $W$ such that  
\begin{subequations}
\label{result}
\begin{align}
&H_{\Omega}(\mf{P}) =  W H_{\Omega}(\mf{R})W^{\dagger}\,,  \label{result1}\\
&\mc{L}_{\gamma}(\mf{P})\vro 
=W \left[\mc{L}_{\gamma}(\mf{R})W^{\dagger} \vro W\right]
W^{\dagger}\,, 
\label{result2}
\end{align}
\end{subequations}
where  
$ W = W_{\mf{u}}(\alpha) $ is given by 
\be
W_{\mf{u}}(\alpha)   =
\exp [ -i \alpha \;\mf{J}^{(1)}\cdot\mf{u}/\hbar ] 
\exp [ -i \alpha \;\mf{J}^{(2)}\cdot\mf{u}/\hbar ] 
 \,. 
\label{definitions}
\ee
Here  the operator 
 $ \exp [ -i \alpha \;\mf{J}^{(\mu)}\cdot\mf{u}/\hbar ]$
describes a  rotation  around the axis $\mf{u}$ by an angle $\alpha$ in the state space of atom $\mu$.
The notation  $H_{\Omega}(\mf{R})$ and 
$\mc{L}_{\gamma}(\mf{R})$  means that the coupling constants and collective decay 
rates in Eqs.~(\ref{H_Omega}) and~(\ref{Lgamma}) have to be evaluated at $\mf{R}$.  

%%%%%%%%%%%%%%%%%%
% proof
%%%%%%%%%%%%%%%%%%
We proceed with the proof of Eq.~(\ref{result}). 
%%%%%%%%%%%%%%%%%%%%%%%%%%%%%%
In a first step, we introduce  the auxiliary operator  
$A_{\mf{R}}=W V_{\mf{R}} W^{\dagger} $,  
where  $V_{\mf{R}}$ is the interaction Hamiltonian 
for a relative position of the atoms given by $\mf{R}$, 
and $W=W_{\mf{u}}(\alpha)$ is  defined in Eq.~(\ref{definitions}). 
The evaluation of $A_{\mf{R}}$ involves only 
the   transformation of  the dipole operator of each atom. Since 
the matrix elements of vector operators transform like  
classical vectors under rotations 
(see, e.g., Sec. 3.10. in~\cite{sakurai:mqm}), we find 
\be 
W\;\mf{\hat{d}}^{(\mu)} \;  W^{\dagger}
=\sum\limits_{i=1}^{3}\big[\mf{\tilde{d}}_{i} \; \Sp{i}{\mu} +\text{H.c.}\big]\,,
\label{rotation}
\ee 
where $\mf{\tilde{d}}_{i} = \rot[-1]{\mf{u}}{\alpha}\mf{d}_i$. 
This shows that the only difference between  the  auxiliary operator  $A_{\mf{R}}$ 
and $V_{\mf{R}}$ is  that the dipole moments of the former are determined 
by  $ \mf{\tilde{d}}_{i}$  instead of $\mf{d}_i$.  
%%%%%%%%%%%%%%%%%%%%%%%%%%%%%%%%%%

In a second step, we employ the tensor properties of $\overset{\leftrightarrow}{\chi}$   
to find the following expression for the parameters $\Omega_{ij}(\mf{P})$ and $\Gamma_{ij}(\mf{P})$
[see Eqs.~(\ref{Omega}) and (\ref{Gamma})], 
\begin{align}
\hbar \Omega_{ij}(\mf{P}) 
 = & \left[\rot[-1]{\mf{u}}{\alpha}  \mf{d}_i\right]^{\text{T}} 
 \TensorRe (\mf{R})  \left[\rot[-1]{\mf{u}}{\alpha} \mf{d}_j^*\right] , \\
 \hbar \Gamma_{ij}(\mf{P}) 
 = & \left[\rot[-1]{\mf{u}}{\alpha} \mf{d}_i\right]^{\text{T}} 
 \TensorIm (\mf{R})  \left[\rot[-1]{\mf{u}}{\alpha} \mf{d}_j^*\right] .
 \label{tensor_omega}  
 \end{align}
This important result shows  that   a rotation of  the dipole moments $\mf{d}_i$  
 by $\rot[-1]{\mf{u}}{\alpha}$ is formally equivalent to  
 a rotation of $\mf{R}$  by $\rot{\mf{u}}{\alpha}$ in the master equation~(\ref{master}). 

 From the combination of the results obtained in step one and two, 
 we conclude that  the exchange of   $V_{\mf{R}}$ by  $A_{\mf{R}}$ 
in the  integral of Eq.~(\ref{master_general})   is equivalent to a rotation of the separation  
vector from $\mf{R}$ to $\mf{P}=\rot{\mf{u}}{\alpha} \mf{R}$, 
\begin{align} 
I  & =
 \frac{-1}{\hbar^2}\int\limits_0^{t}
d\tau \text{Tr}_{\text{F}}\Big(\big[A_{\mf{R}}  ,  
\big[U(\tau) A_{\mf{R}}U^{\dagger}(\tau),  \tilde{\vro}(\hat{\tau}) \big]\big]\Big)
\label{integral} \\ 
 & =  -\frac{i}{\hbar} [H_{\Omega}(\mf{P}) , \vro] +     \mc{L}_{\gamma}(\mf{P})\vro\,,
\label{step1}
\end{align}
where  $\hat{\tau}=t-\tau$ and 
\be 
\tilde{\vro}(\hat{\tau}) = U(\tau)\left[\vro_{\text{F}}\otimes\vro(\hat{\tau})\right]U^{\dagger}(\tau) \,.
\label{tilde_rho}
\ee
%%%%%%%%%%%%%%%%%%%%%%
Note that the equality of Eqs.~(\ref{integral}) and~(\ref{step1}) holds 
under the same assumptions that led from Eqs.~(\ref{master_general}) to~(\ref{master}).  
%%%%%%%%%%%%%%%%%%%%%%%

In the second part of the proof 
we evaluate the integral in Eq.~(\ref{integral}) in a different way.  
In the discussion following Eq.~(\ref{chi}),  we justified that   
$\mc{L}_{\gamma}$ and $H_{\Omega}$  depend only on the 
mean transition frequency $\omega_0$.  
Here we employ exactly the same approximation~\cite{remark:1}   
and replace the frequencies $\omega_i$ appearing in $U(\tau) A_{\mf{R}}U^{\dagger}(\tau)$ by $\omega_0$.  
Since  $H_{\text{A}}$ commutes with $\mf{J}^{(\mu)}$  
if all frequencies $\omega_i$ are replaced by the mean transition frequency 
$\omega_0$, we have  $[W,U]=0$ and hence 
\be 
U(\tau) A_{\mf{R}}U^{\dagger}(\tau) 
= W U(\tau)  V_{\mf{R}} U^{\dagger}(\tau) W^{\dagger}\,. 
\ee  
It follows that  the argument of the trace in Eq.~(\ref{integral}) can be written as 
\be
W\big[V_{\mf{R}}, \,
\big[U(\tau) V_{\mf{R}}U^{\dagger}(\tau) , W^{\dagger} \tilde{\vro}(\hat{\tau}) W \big] \big] W^{\dagger}\,.
\ee
In contrast to Eq.~(\ref{integral}), the double commutator contains now the original interaction 
Hamiltonian $V_{\mf{R}}$ that corresponds  to a setting with separation vector $\mf{R}$. We thus obtain
\be
I =    -\frac{i}{\hbar} [W H_{\Omega}(\mf{R}) W^{\dagger} , \vro] 
  +   W\left[\mc{L}_{\gamma}(\mf{R}) W^{\dagger}\vro W \right] W^{\dagger}\,.
\label{second_step}
\ee
Finally, the comparison of Eqs.~(\ref{second_step}) and~(\ref{step1})  establishes  Eq.~(\ref{result})  
which concludes the proof. 

Note that throughout this proof, we have not made reference to
the specific type of the Zeeman sublevels employed in our example
shown in Fig.~\ref{picture1}. Therefore, the central theorem holds
for transitions between states with arbitrary angular momentum 
structure, as long as complete multiplets are considered.

\subsection{\label{sec:diag-1}Diagonalization of $H_{\Omega}$}
%%%%%%%%%%%%%%%%%%%%
%   discussion
%%%%%%%%%%%%%%%%%%%%
We now turn to the discussion of   Eq.~(\ref{result}),
which will lead to our central results.  
The Hamiltonian $H_{\Omega}$ describes 
the coherent part of the dipole-dipole interaction between the atoms.
From  Eq.~(\ref{result1}), it is immediately clear that the eigenvalues of  
$H_{\Omega}$  depend only on the interatomic distance, but not on  
the  orientation of the separation vector $\mf{R}$.  The reason is that
the spectrum of two operators, which are related by a unitary 
transformation, is identical. In our case, the Hamiltonian
$H_{\Omega}(\mf{R})$ and $H_{\Omega}(\mf{P})$ for different orientations
$\mf{R}$ and $\mf{P}$ are related by the unitary transformation $W$,
and since $\mf{P}$ is obtained from $\mf{R}$ by an arbitrary rotation, 
the eigenvalues of $H_{\Omega}$ are identical for any orientation.
%%%%%

Next we  re-obtain this result  in a more explicit way  
and  derive symbolic expressions for the eigenvalues 
and eigenstates of $H_{\Omega}$.  
This Hamiltonian  can be written as 
\be
H_{\Omega}   =   \sum\limits_{i,j=1}^3\big( \;[H_{\Omega}]^{\mc{S}}_{ij}\;\ket{s_i}\bra{s_j} 
+ [H_{\Omega}]^{\mc{A}}_{ij} \;\ket{a_i}\bra{a_j} \;\big)  \,,
\label{expansion_HO}
\ee
where the symmetric and antisymmetric states are defined as 
\begin{subequations}
\label{states}
\begin{align}
\ket{s_i}   &=   (\,\ket{i,4}+\ket{4,i}\,)/\sqrt{2} \,,\\
\ket{a_i}   &=   (\,\ket{i,4}-\ket{4,i}\,)/\sqrt{2}\,,
\end{align}
\end{subequations}
and $\ket{i,j} = \ket{i_1}\otimes\ket{j_2}$.  
Since all matrix elements $\bra{s_i}H_{\Omega}\ket{a_j}$ 
of $H_{\Omega}$ between a symmetric and an antisymmetric state 
vanish,  the set of eigenstates   decomposes into
a symmetric subspace $\mc{S}$ and an antisymmetric subspace $\mc{A}$.
The matrix elements of $H_{\Omega}$ in the subspace $\mc{S}$ spanned by the symmetric states 
\mbox{$\{\ket{s_1},\,\ket{s_2},\,\ket{s_3} \}$}  are 
\begin{align}
[H_{\Omega}]^{\mc{S}} & = -\hbar 
\left(
\begin{array}{r@{\hspace{0.5cm}}r@{\hspace{0.5cm}}r }
\Omega_{11}      & \Omega_{21}^*            &  \Omega_{31}^*   \\
\Omega_{21}     & \Omega_{22} & \Omega_{32}^*  \\
\Omega_{31} & \Omega_{32} & \Omega_{33} 
\end{array}
\right)\,,
\label{H_Omega_S}
\end{align}
and the representation of  $H_{\Omega}$ in the subspace $\mc{A}$ spanned by 
the   antisymmetric states 
\mbox{$\{\ket{a_1},\,\ket{a_2},\,\ket{a_3} \}$} is given by 
$[H_{\Omega}]^{\mc{A}}= - [H_{\Omega}]^{\mc{S}}$. 
Note that the collective ground state $\ket{4,4}$ and the   states 
$\ket{i,j}$  ($i,j \in\{1,2,3\}$) where each atom is in an 
excited state are not influenced by the dipole-dipole interaction and thus not 
part of the expansion~(\ref{expansion_HO}). 

In Section~\ref{sec:theorem}, we have derived a general relation
between any two orientations of the interatomic distance vector.
In order to apply this result, we define the vector $\mf{R}_z$ to be 
parallel to the $z$ axis, i.e.  $\mf{R}_z= R\, \mf{e}_z$. 
This corresponds to the choice $\theta = 0$ in Eq.~(\ref{coord}).
Any separation vector $\mf{P}$ can then be obtained from
$\mf{R}_z$ as $\mf{P}=\rot{\mf{u}}{\alpha} \mf{R}_z$
by a suitable choice of the rotation 
axis $\mf{u}$ and the angle $\alpha$.

We then proceed with the diagonalization of 
the Hamiltonian $H_{\Omega}(\mf{R}_z)$
with atomic separation vector $\mf{R}_z$.
The explicit calculation of the coupling constants $\Omega_{i j}$
shows that  the off-diagonal elements in Eq.~(\ref{H_Omega_S}) 
vanish if the atoms are aligned along the $z$ axis,
see Eqs.~(\ref{OmegaExplicit}) and (\ref{GammaExplicit}) with $\theta = 0$. 
It follows  that the Hamiltonian $H_{\Omega}(\mf{R}_z)$ is already diagonalized 
by the symmetric and antisymmetric states Eq.~(\ref{states}), 
and the eigenvalues of  $[H_{\Omega}]^{\mc{S}}$ and $[H_{\Omega}]^{\mc{A}}$ 
are given by  
$\lambda^{\mc{S}}_i = -\hbar \Omega_{ii}(\mf{R}_z)$ and 
$\lambda^{\mc{A}}_i = \hbar \Omega_{ii}(\mf{R}_z)$, respectively.

According to Eq.~(\ref{result1}), 
the  Hamiltonian $H_{\Omega}(\mf{P})$ is the unitary transform of 
$H_{\Omega}(\mf{R}_z)$ by  $W$. 
The  normalized eigenstates 
of $H_{\Omega}(\mf{P})$ are thus determined by  $W\ket{s_i}$  
and $W\ket{a_i}$, and  their eigenvalues are again 
$\lambda^{\mc{S}}_i$ and $\lambda^{\mc{A}}_i$ , respectively. 
Since  the orientation of $\mf{P}$ is arbitrary,  
the eigenvalues of $H_{\Omega}(\mf{P})$ depend only on 
the interatomic distance  $|\mf{P}|=|\mf{R}_z|=R$, but not on 
the orientation of the separation vector.  

Thus, it follows from our theorem in Sec.~\ref{sec:theorem} that 
the eigenvalues of $H_{\Omega}(\mf{P})$ are invariant under rotation
of the interatomic distance vector. 
%%%%%%%%%%%%%%%%%%%%%%

\subsection{\label{sec:diag-2}Diagonalization of $H_{\text{A}} +H_{\Omega}$}
An additional conclusion can be drawn from Eq.~(\ref{result}) 
if the operator $H_{\text{A}}$ commutes with the transformation 
$W= W_{\mf{u}}(\alpha)$, i.e., 
\begin{equation}
[H_{\text{A}},W]=0\,. 
\end{equation}
Then, Eq.~(\ref{result1}) implies that 
 $H_{\text{A}} +H_{\Omega}(\mf{P})$  is 
the unitary transform of  $H_{\text{A}} +H_{\Omega}(\mf{R})$ by  $W$. 
A straightforward realization of this is the case of vanishing
Zeeman splitting $\delta$,  in which the relation holds for an 
arbitrary orientation of $\mf{P}$.
Then, the energy levels of the full system Hamiltonian $H_{\text{A}} +H_{\Omega}$ 
do not depend on the  orientation of the separation vector. 

This result can be understood as follows.  
In the absence of a magnetic field ($\delta=0$), there is no distinguished direction in space. 
Since the vacuum is isotropic in free space, one expects  that the energy levels  
of the  system  are invariant under rotations of the separation vector $\mf{R}$.

By contrast, the application of a magnetic field in $z$ direction 
breaks the full rotational symmetry. 
For $\delta\not=0$, the atomic Hamiltonian $H_{\text{A}}$ 
only commutes with   transformations $W_{\mf{u}}(\alpha)$ 
that correspond to a rotation of the separation vector 
around the $z$ axis, $\mf{u}=\mf{e}_z$.
If we express the atomic separation vector in terms of spherical coordinates 
as in Eq.~({\ref{coord}), this means that the  eigenvalues of the 
full system Hamiltonian $H_{\text{A}} +H_{\Omega}$ do only depend 
on the interatomic distance $R$ and the angle $\theta$, but not  on 
the angle $\phi$.    
This result reflects the symmetry of   our system with respect to 
rotations around the $z$ axis.

\subsection{\label{sec:evolution}Unitary equivalence of time evolution in different orientations}
If the operator $H_{\text{A}}$ commutes with the transformation 
$W= W_{\mf{u}}(\alpha)$, another conclusion can be drawn.  
Then,  the result in Eq.~(\ref{result}) implies that the density operator $W \vro(\mf{R}) W^{\dagger}$ 
obeys the same master equation than $\vro(\mf{P})$  
for $\mf{P}=\rot{\mf{u}}{\alpha} \mf{R}$. It follows that $\mf{P}$ is the  unitary transform of 
$\vro(\mf{R})$ by $W$, i.e. 
\begin{equation}
\vro(\mf{P}) =W \vro(\mf{R}) W^{\dagger} \,.
\end{equation}
As discussed in Sec.~\ref{sec:diag-2}, 
the free atomic Hamiltonian $H_{\text{A}}$   commutes with  
$W_{\mf{u}}(\alpha)$ for an arbitrary choice 
of the rotation axis $\mf{u}$ and angle $\alpha$
if the Zeeman splitting $\delta$ 
vanishes.  

We thus conclude that it suffices to 
determine the solution of the master equation~(\ref{master}) for only one 
particular geometry if $\delta=0$. %
Any other solution  can then be generated simply by applying 
the transformation $W= W_{\mf{u}}(\alpha)$ with suitable values 
of $\mf{u}$ and $\alpha$ to the solution for the particular geometry. 

\subsection{\label{sec:establishment}Establishment of the breakdown} 
%The breakdown of the few-level approximation for collective systems
%is established by noting that the result in  Eq.~(\ref{result}) and all its implications 
%in Secs.~\ref{sec:diag-1}-\ref{sec:evolution}
%cannot be recovered if  any of the Zeeman sublevels of the $P_1$ triplet are neglected.  
%In this case, the unitary operator $W$ does not exist since  
%it is impossible to define an angular momentum or vector operator 
%in a  state space where magnetic sublevels have been removed artificially. 
%In particular, an artificially reduced sublevel scheme will exhibit eigenenergies
%with a spurious dependence on the orientation of the interatomic distance vector,
%in contrast to our findings in Secs.~\ref{sec:diag-1} and \ref{sec:diag-2}.
%%%%%%%%
In Secs.~\ref{sec:diag-1}-\ref{sec:evolution}, 
we presented several  results concerning  the energy levels and the time evolution 
of the two dipole-dipole interacting atoms that 
are based on the central theorem in Eq.~(\ref{result}).  
However, this theorem can only be established if each atom is modelled by 
complete sets of angular momentum multiplets, and represents
the reference case that corresponds to results which can be 
expected in an experiment. 
If  any of the Zeeman sublevels of the $P_1$ triplet are neglected,  
the unitary operator $W$ does not exist since  
it is impossible to define an angular momentum or vector operator 
in a  state space where magnetic sublevels have been removed artificially. 
In this case, the central statement cannot be applied. Still,
the system can be solved without the help of the theorem.
The breakdown of the few-level approximation for collective systems
is then established by noting that the results for systems with 
artificially reduced state space fail to recover the results
derived in Secs.~\ref{sec:diag-1}-\ref{sec:evolution} for the full
system.

%f any of 
%the Zeeman sublevels of the $P_1$ triplet are omitted by a direct 
%calculation. 

In order to illustrate this point in more detail,  
we consider the system in Fig.~\ref{picture1}  and 
assume that the excited states  of each atom are degenerate ($\delta=0$). 
According to our findings in Secs.~\ref{sec:diag-1}  and \ref{sec:diag-2}, 
the energy levels of the complete system depend on the length of the 
separation vector $\mf{R}$, but not on its orientation.  
In contrast, the omission of any of the Zeeman sublevels leads to 
a spurious dependence of the energy levels on the orientation, 
and thus to incorrect predictions. 

For example, if the excited states $\ket{1}$ and $\ket{3}$ in each atom 
are omitted, the level scheme in Fig.~\ref{picture1}(b) reduces to 
an effective two-level system comprised of the states $\ket{2}$ 
and $\ket{4}$.  The collective two-atom system is then described 
by the ground state $\ket{4,4}$, the excited state $\ket{2,2}$ 
and the symmetric and antisymmetric states $\ket{s_2}$ 
and $\ket{a_2}$.   The frequency splitting between the  
states $\ket{s_2}$ and $\ket{a_2}$ is given by $2|\Omega_{22}|$, 
where 
\be
\Omega_{22}  =    \frac{3}{2}
\gamma_2 \left[ f_1(\eta) - \cos^2(\theta) f_2(\eta) \right]\,,
\label{twolevel_splitting}
\ee
and 
\begin{align}
f_1(\eta) & = \left(\frac{1}{\eta} -\frac{1}{\eta^3}\right) \cos\eta 
-\frac{1}{\eta^2}\sin\eta  \,, \\
f_2(\eta) & = \left(\frac{1}{\eta} - \frac{3}{\eta^3}\right)\cos\eta -\frac{3}{\eta^2}\sin\eta  \,.  
\end{align} 
Since the second term in Eq.~(\ref{twolevel_splitting}) is proportional 
to $\cos^2\theta$, the energy levels of the artificially created two-level 
system strongly depend on the orientation of the separation vector $\mf{R}$. 
This is at variance with our finding in Sec.~\ref{sec:diag-2}, where 
we have shown that the energy levels 
do not depend on the orientation of the vector $\mf{R}$ if 
each atom consists of complete and degenerate Zeeman multiplets. 
We thus conclude that all Zeeman sublevels generally have to be taken into account. 
%%%%%%%

Since the validity of the central theorem 
Eq.~(\ref{result}) is not restricted to the $S_0\leftrightarrow P_1$ transition
discussed so far,
it follows that the breakdown of the few-level approximation
can be established for transitions between arbitrary angular 
momentum multiplets.

The intuitive explanation of the breakdown has already been hinted at in Sec.~\ref{sec:example}.
For a more formal discussion, we return to the matrix representation of 
$[H_{\Omega}]^{\mc{S}}$ in Eq.~(\ref{H_Omega_S}). 
The diagonal elements proportional to $\Omega_{ii}$ account for the  coherent 
interaction between a dipole of one of the atoms and the corresponding dipole of the other atom. 
By contrast, the off-diagonal terms  proportional to $\Omega_{ij}$ with $i\not=j$ arise from 
the vacuum-mediated    interaction between orthogonal 
dipoles of different atoms~\cite{agarwal:01,evers:06}. 
It is the presence of these terms that renders the simplification of the 
atomic level scheme impossible since they 
couple an excited state $\ket{i}$ of one atom to a different 
excited state $\ket{j}$ ($i\not=j$) of the other atom. 
A similar argument applies to the collective decay  rates $\Gamma_{ij}$  
appearing in $\mc{L}_{\gamma}\vro$. 
Thus, if any Zeeman sublevel of the excited state multiplet is artificially removed, then 
some of these vacuum-induced couplings $\Omega_{ij}$ with $i\not=j$ are neglected,
which leads to incorrect results. Now, it is also apparent why the breakdown of the 
few-level approximation appears exclusively in collective systems. For single atoms
in free space, a coupling of orthogonal transition dipole moments via the
vacuum is impossible. 

\subsection{\label{sec:recovery}Recovery of the few-level approximation in special geometries}
The identification of the vacuum-induced couplings $\Omega_{ij}$ and $\Gamma_{ij}$  between 
orthogonal transition dipole moments as the cause of the breakdown 
enables one to conjecture that few-level approximations 
are justified for particular geometrical setups, where some or all of 
the cross-coupling terms 
vanish. 

For example, we mentioned earlier that all cross-coupling terms vanish if 
the atoms are aligned along the $z$ axis. This corresponds to the case $\theta=0$ in 
Eqs.~(\ref{coord})-(\ref{GammaExplicit}).
Then, the $S_0\leftrightarrow P_1$ transition 
may be reduced to a two-level system, formed by an arbitrary sublevel of the $P_1$ 
triplet and the ground state $S_0$. 

As a second example, we assume the 
atoms to be aligned in the $x$-$y$-plane, i.e., $\theta=\pi/2$ in 
Eq.~(\ref{coord}). Then the 
terms $\Omega_{21}$, $\Gamma_{21}$ and $\Omega_{32}$, $\Gamma_{32}$ vanish,
see Eqs.~(\ref{OmegaExplicit})-(\ref{GammaExplicit}). 
In effect, the excited state $\ket{2}$ 
may be disregarded such that the atomic level scheme  simplifies to a V-system 
formed by the states $\ket{1}$ and $\ket{3}$ of the $P_1$ multiplet and the ground 
state $S_0$. 

Note that the cross-coupling terms also become irrelevant
in the special case  
$|\delta| \gg |\Omega_{ij}|, |\Gamma_{ij}|$ ($i\neq j$),
see our discussion below Eq.~(\ref{Gamma}).

\section{\label{sec:summary}Discussion and summary}
Throughout this article, we have studied the properties of various parts
of the system Hamiltonian as well as the full density operator under rotations
of the interatomic distance vector. This discussion was based on a general theorem
in Sec.~\ref{sec:theorem} which relates the system properties for different
orientations of the interatomic distance vector. 

First, we have discussed the Hamiltonian $H_{\Omega}$, which describes the coherent
coupling between different transitions in the two atoms induced by the vacuum field.
Armed with our main theorem, it is possible to first diagonalize $H_{\Omega}$
in a special geometry, where the eigenvectors and eigenenergies assume a particularly
simple form. The eigenvectors and eigenenergies for an arbitrary system geometry
are then derived via the theorem. Our main result of Sec.~\ref{sec:diag-1} is that 
the eigenvalues of $H_{\Omega}$ are invariant under rotation
of the interatomic distance vector.

In a second step, we have studied the eigenenergies of the full system 
Hamiltonian $H_{A}+H_{\Omega}$,
which in general are {\it not} invariant under rotation
of the interatomic distance vector. The invariance, however, is recovered if
$H_{\text{A}}$ commutes with the transformation $W= W_{\mf{u}}(\alpha)$,
which is given in explicit form as a result of our theorem.
Most importantly, this additional condition is fulfilled for a degenerate 
excited state multiplet, i.e., if the Zeeman splitting $\delta$ vanishes.
Then, there is no preferred direction in space, such that the invariance of
the eigenenergies, which are observables, can be expected.

We then conclude the breakdown of the few-level approximation in Sec.~\ref{sec:establishment}, 
since our results of the previous sections are violated if any of the excited
state multiplet sublevels are artificially removed. Possible consequences are,
for example, a spurious dependence of the eigenenergies on the orientation of 
the interatomic distance vector, and thus of all observables that depend
on the transition frequencies among the various eigenstates of the system. 
In experiments, in addition, a loss of population from the subspace 
considered in the few-level approximation would be observed.
Our proof can be generalized to
transitions between arbitrary angular momentum multiplets.

We have identified the vacuum-induced dipole-dipole coupling between transitions with
orthogonal dipole moments as the origin of the breakdown. On the one hand, this explains
why the breakdown exclusively occurs in collective systems, since such orthogonal
couplings are impossible in single atoms in free space. On the other hand, 
the interpretation enables one to identify special geometries where some of the
Zeeman sublevels can be omitted. This also allows to connect our results to previous
studies involving dipole-dipole interacting few-level systems. 
In these studies
involving the few-level approximation, typically a very special geometry was chosen, e.g., 
with atomic separation vector and transition dipole moments orthogonal or parallel to 
each other. These results remain valid if a geometry can be found such that 
the full Zeeman sublevel scheme reduces to the chosen level scheme 
as discussed in Sec.~\ref{sec:recovery}. It should be noted, however, that
there are physical realizations of interest which in general do not allow 
for a particular system geometry that leads to the validity of a few-level approximation,
such as quantum gases.

Finally, on a more technical side, our results can also be applied to 
considerably simplify the computational effort required for the 
treatment of such dipole-dipole interacting multilevel systems with arbitrary
alignment of the two atoms.
First, our theorem both allows for a convenient evaluation of eigenvalues and
eigenenergies for arbitrary orientations of the interatomic distance
vector based on the results found in a single, special alignment.
Second, we have found in Sec.~\ref{sec:evolution} that for the degenerate
system, the density matrices for different orientations are related to
each other by the unitary transformation $W$ defined in our theorem. 
Thus the solution for any orientation can be obtained from a single
time integration simply by applying this transformation.


\begin{thebibliography}{21}
\bibitem[{rem()}]{remark:1}
\bibinfo{note}{Note that $H_A$ appearing in the free evolution term
  $-i[H_A,\vro]/\hbar$ of the master equation is not approximated.}
\end{thebibliography}
\end{document}